\documentclass[pdflatex,sn-mathphys-ay]{sn-jnl}
\PassOptionsToPackage{hyphens}{url}

\usepackage[utf8]{inputenc} 
\usepackage[T1]{fontenc}    
\usepackage{graphicx}
\usepackage{booktabs}
\providecommand{\burl}[1]{\url{#1}}
\usepackage{amsfonts}       
\usepackage{nicefrac}       
\usepackage{xcolor}         
\hypersetup{pdfauthor={Anonymous}}
\usepackage{tikz}
\usetikzlibrary{positioning, arrows.meta, shapes.geometric}
\usetikzlibrary{shadows.blur}

\usepackage[acronym]{glossaries}
\glsdisablehyper
\newacronym{llm}{LLM}{large language model}
\newacronym{mas}{MAS}{multi-agent systems}
\newacronym{tr}{TR}{task recognition}
\newacronym{tl}{TL}{task learning}
\newacronym{rise}{RISE}{Recursive Introspection}
\newacronym{rfm}{RFM}{Relational Forward Models}
\newacronym{smart}{SMART}{Smart Multi-Agent Robot Task Planning}
\newacronym{marl}{MARL}{multi-agent reinforcement learning}
\newacronym{gabm}{GABM}{generative agent-based modeling}
\newacronym{cgcot}{CGCoT}{Concept-Guided Chain-of-Thought}

\tikzset{
    mainbox/.style = {
        rectangle, rounded corners, draw=black, very thick,
        fill=blue!10,
        minimum width=5cm, minimum height=1.2cm,
        align=center, font=\bfseries\Large
    },
    category/.style = {
        rectangle, rounded corners, draw=black, thick,
        fill=cyan!10,
        minimum width=4.5cm, minimum height=1cm,
        align=center, font=\bfseries
    },
    method/.style = {
        rectangle, draw=black, thick,
        fill=gray!10,
        minimum width=5cm, minimum height=0.9cm,
        align=center, font=\small\itshape
    },
    smallbox/.style = {
        rectangle, draw=black, thick,
        fill=gray!5,
        minimum width=3.5cm, minimum height=0.8cm,
        align=center, font=\small
    },
    arrow/.style = {
        thick, -{Stealth[length=3mm, width=2.5mm]}, draw=black
    },
    node distance=1.2cm and 1.8cm
} 

\makeatletter
\def\toclevel@subparagraph{2}
\def\Hy@toclevel@subparagraph{2}
\def\toclevel@bmhead{2}
\def\Hy@toclevel@bmhead{2}
\makeatother

\title[Open-Ended Co-Evolution in LLM-MAS]{Static Sandboxes Are Inadequate: Modeling Societal Complexity Requires Open-Ended Co-Evolution in LLM-Based Multi-Agent Simulations}

\author*[1]{\fnm{Jinkun} \sur{Chen}}\email{jinkun.chen@dal.ca}
\author[1]{\fnm{Sher} \sur{Badshah}}\email{sh545346@dal.ca}
\author[1]{\fnm{Xuemin} \sur{Yu}}\email{xuemin.yu@dal.ca}
\author[2]{\fnm{Sijia} \sur{Han}}\email{hansijia@meta.com}

\affil[1]{\orgdiv{Faculty of Computer Science}, \orgname{Dalhousie University}, \orgaddress{\city{Halifax}, \country{Canada}}}
\affil[2]{\orgname{Meta}, \orgaddress{\city{Vancouver}, \country{Canada}}}

\begin{document}

\abstract{What if artificial agents could not just communicate, but also evolve, adapt, and reshape their worlds in ways we cannot fully predict? With \glspl{llm} now powering multi-agent systems and social simulations, we are witnessing new possibilities for modeling open-ended, ever-changing environments. Yet, most current simulations remain constrained within static sandboxes, characterized by predefined tasks, limited dynamics, and rigid evaluation criteria. These limitations prevent them from capturing the complexity of real-world societies. In this paper, \textbf{we argue that static, task-specific benchmarks are fundamentally inadequate and must be rethought.} We critically review emerging architectures that blend \glspl{llm} with multi-agent dynamics, highlight key hurdles such as balancing stability and diversity, evaluating unexpected behaviors, and scaling to greater complexity, and introduce a fresh taxonomy for this rapidly evolving field. Finally, we present a research roadmap centered on open-endedness, continuous co-evolution, and the development of resilient, socially aligned AI ecosystems. \textbf{We call on the community to move beyond static paradigms and help shape the next generation of adaptive, socially-aware multi-agent simulations.}}


\maketitle

\begin{figure}[htbp]
    \centering
    \includegraphics[width=0.9\linewidth]{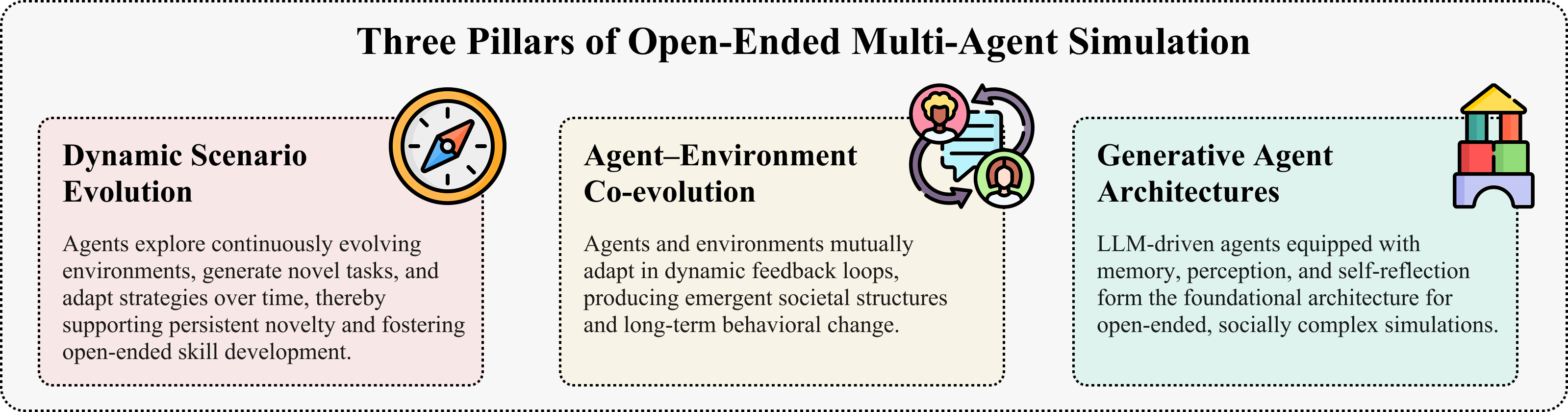}
    \caption{Our proposed taxonomy of open-ended multi-agent simulation: (1) \textit{Dynamic Scenario Evolution}, (2) \textit{Agent–Environment Co-evolution}, and (3) \textit{Generative Agent Architectures}. These pillars support adaptive, socially aligned LLM-driven ecosystems.}
    \label{fig:pillars}
\end{figure}

\section{Introduction}

Today's multi-agent simulations remain largely confined within static sandboxes: agents are evaluated on predefined tasks, in closed environments, using narrow metrics~\citep{liu_agentbench_2023,yehudai_survey_2025,luo_large_2025,lin_agentsims_2023,jin_llms_2025}. This rigidity stifles emergence, suppresses innovation, and limits the potential of truly adaptive agents. Treating adaptability, societal feedback, and long-term transformation as afterthoughts is no longer acceptable in an era of increasingly capable AI. Despite recent progress, the field remains fragmented across diverse methodologies, architectures, and evaluation standards. We highlight that without a unified perspective grounded in open-endedness, researchers risk optimizing brittle systems that cannot generalize beyond toy domains. \textbf{We argue that current paradigms of multi-agent simulation, which are grounded in static tasks, rigid benchmarks, and predictable interaction loops, are fundamentally inadequate for modeling the complexity of real-world societies.} As \glspl{llm} increasingly power \gls{mas}~\citep{NEURIPS2024_ca9567d8}, a new opportunity emerges: to create dynamic, adaptive, and open-ended ecosystems where agents co-evolve with their environments and with each other. 

Imagine a digital society where intelligent agents do not just complete tasks, but evolve their own cultures, languages, and societal structures. These agents adapt to unexpected events, reshaping their environments and forging new collective behaviors in ways beyond researchers' predictions. \textbf{We believe that such open-ended co-evolution, rather than fixed performance, represents the true frontier for adaptive AI.} \textbf{We take a principled stance: multi-agent simulations must shift from task-specific optimization toward modeling the open-ended, co-evolutionary dynamics of real societies.} In addition, 
we contend that unpredictability is not a flaw to be controlled but a feature to be embraced. This view is supported by recent research in multi-agent systems, where unpredictability enables more complex emergent behaviors that are integral to multi-agent systems~\citep{guo_large_2024} and creates opportunities for novel cooperative behaviors to emerge~\citep{lu_generative_2024}, demonstrating that complexity and unpredictability serve as valuable characteristics rather than problems to eliminate.
Only by shifting priorities toward open-ended co-evolution can we build LLM-driven simulations that reflect the richness, resilience, and complexity of real-world societies. This approach allows us to move past the brittle constraints of static sandbox environments.


To support this position, we begin by redefining key constructs such as \glspl{llm}, multi-agent systems, and social simulation through the lens of open-endedness. We then examine the systemic limitations of static simulation paradigms and argue for a conceptual shift toward dynamic, co-evolutionary systems. Building on this foundation, we analyze emerging frameworks that integrate \glspl{llm} into multi-agent architectures and explore their capacity for reasoning, communication, and norm emergence. As a core contribution, we define a conceptual taxonomy that is structured around three foundational pillars: \textit{\textbf{Dynamic Scenario Evolution}}, \textit{\textbf{Agent–Environment Co-evolution}}, and \textit{\textbf{Generative Agent Architectures}} (see~\autoref{fig:pillars}). This taxonomy is not merely descriptive, which reflects our call to action: to design simulations where agents not only perform but also evolve, not only act but also adapt, not only learn but also transform. Finally, we outline emerging applications, evaluation challenges, and a roadmap for advancing adaptive, socially aligned simulation ecosystems.

\section{Background and Definitions}
Before we can reimagine simulation design, it is imperative to clarify its conceptual foundations. In this section, we revisit core constructs such as \glspl{llm}, \gls{mas}, generative agents, and social simulation. These are not presented as neutral definitions; rather, they reflect underlying philosophical commitments. Each represents a fork in the road between static modeling and dynamic co-evolution. We assert that redefining these terms through the lens of open-endedness is critical for realigning the field.

\subsection{Reframing Core Constructs}

\subsubsection{LLMs as Adaptive Cognitive Engines} \Glspl{llm} are deep neural networks trained on massive corpora to predict the next token in a sequence~\citep{NEURIPS2020_1457c0d6}. While traditionally viewed as text generators, we argue they are better understood as adaptive cognitive engines—entities capable of performing complex reasoning, in-context learning, and flexible communication~\citep{bubeck2023sparksartificialgeneralintelligence}. Their real power lies in emergent, context-sensitive behavior that allows for dynamic participation in evolving environments rather than static task completion. As such, they offer a foundation for developing agents that can self-reflect, revise their beliefs, and participate in socially grounded simulations.

\subsubsection{MAS Beyond Coordination: Toward Norm Fluidity} \Gls{mas} consist of autonomous entities interacting within a shared environment to achieve individual or collective goals~\citep{Wooldridge_Jennings_1995}. Classic \gls{mas} research emphasizes coordination under fixed rules and goals~\citep{10.55551483085}. With \gls{llm} integration, however, agents gain the capacity for real-time negotiation, role adaptation, and the emergence of novel social norms~\citep{NEURIPS2024_fa54b0ed}. This transforms \gls{mas} from tools for optimization into frameworks for exploring how agents evolve social identities, institutions, and values through continuous interaction.

\subsubsection{Generative Agents as Normative Actors} Generative agents are LLM-driven entities equipped with memory, perception, and reflection capabilities~\citep{yao2023react}. They simulate lifelike social behavior and cognitive processes~\citep{10114535861833606763}. Moving beyond scripted responses, these agents engage in belief revision, identity formation, and social norm negotiation~\citep{cheng2024exploringlargelanguagemodel, hua-etal-2024-assistive, jiao_navigating_2025}. Their internal architectures enable recursive introspection and social reasoning, positioning them as experimental probes for understanding norm emergence, trust dynamics, and social contract formation.

\subsubsection{Social Simulation as Emergence, Not Reproduction} It refers to the computational modeling of social phenomena, traditionally implemented through rule-based, agent-based models~\citep{gilbert2005simulation}. Early models aimed to replicate known dynamics (e.g., economic markets or crowd behavior). In contrast, LLM-enabled simulations facilitate the emergence of novel cultural patterns, institutions, and behaviors~\citep{siebers_exploring_2024, dizaji_incentives_2024}. These agents not only adapt to existing rules but invent new interaction protocols, offering an unprecedented platform for simulating the evolution of social complexity across different environments and timescales.

\subsection{Rethinking Simulation Paradigms}

Common challenges in LLM-MAS: hallucination, memory drift, value misalignment, and scalability, are often viewed as engineering flaws. Yet these `failures' frequently reveal hidden assumptions in system design. For instance, hallucination can expose insufficient grounding mechanisms~\citep{pan_what_2023}; memory issues reveal the inadequacy of short-context reasoning~\citep{hou_my_2024}; and cultural bias illustrates the tension between universal models and local values~\citep{kamruzzaman_woman_2024}. Rather than suppress these behaviors, we propose treating them as entry points for open-ended inquiry. Each failure mode highlights a mismatch between static assumptions and dynamic system needs—a mismatch that future research should explore, not erase.

Current research primarily continues to assume relatively static environments and predictable interaction patterns, an assumption increasingly misaligned with the demands of simulating real-world societies. While recent progress in generative agents, norm fluidity, and emergent simulation is encouraging~\citep{mou2024individual}, much of the work still operates within narrowly scoped, task-centric paradigms~\citep{larooij2025large}. These models often reinforce brittle dynamics and fail to account for the recursive nature of social adaptation, where agents influence their environment and are simultaneously reshaped by it.

\textbf{We argue that this conservative framing limits not only the scope of possible agent behaviors but also the relevance of simulation outcomes to real-world complexity.} Critical phenomena such as value renegotiation, institutional breakdown, coalition formation, and cultural drift are rarely modeled, and when they are, it's typically within a constrained rule space. A deeper understanding of how agents respond to rapid, discontinuous changes and emergent non-linear dynamics remains underexplored and undervalued.

\section{Integrating LLMs into Multi-Agent Systems}\label{sec:llm-mas}
We contend that current LLM-MAS integrations overwhelmingly prioritize performance and predictability, often at the cost of adaptability, norm emergence, and long-term societal co-evolution. \textbf{We argue that to unlock the full potential of \glspl{llm} in multi-agent systems, future architectures must emphasize intrinsic mechanisms for dynamic goal realignment, identity fluidity, and collaborative reasoning under open-ended conditions.}

This section critically examines emerging frameworks, reasoning methods, and communication paradigms, while highlighting challenges and future directions for building scalable, interpretable, and socially aligned LLM-integrated \gls{mas}. See Table~\ref{tab:frameworks} in Appendix~\ref{appendix-a} for a detailed overview of key frameworks. In distributed \gls{mas} architectures, on-device \gls{llm} deployment enables collaborative problem-solving while minimizing communication overhead~\citep{zou_wireless_2023}. Resource-efficient frameworks such as nanoLM~\citep{yao_nanolm_2024} further facilitate scaling by predicting performance without the cost of full-model training. Applications in economic simulations, like AI-Economist~\citep{dizaji_incentives_2024}, showcase the versatility of \glspl{llm} in modeling complex socio-economic behaviors.

\subsection{Reasoning and Decision-Making}
\glspl{llm} extend agent reasoning well beyond reactive policies by equipping agents with contextual planning capabilities through techniques such as iterative prompting~\citep{wang_voyager_2023}, which structures multi-step deliberation, and tagged context prompts~\citep{feldman_trapping_2023} that mitigate spurious outputs. Employing \glspl{llm} in modeling social influence and private information dynamics, the TwinMarket~\citep{yang_twinmarket_2025} multi-agent framework especially illustrates the nuanced reasoning capacities unlocked by \glspl{llm}. 

Moreover, architectural innovations in LLM-MAS integration, including Dynamic Diffusion~\citep{zhang_understanding_2025}, Venn Diagram Prompting~\citep{mahendru_venn_2024}, and Grammar Masking~\citep{netz_using_2024}, are able to refine agent decision-making by prioritizing relevant information and ensuring logical consistency. Meanwhile, frameworks like UniBias~\citep{zhou_unibias_2024} and long-term memory modules~\citep{hou_my_2024} offer promising strategies for aligning agent reasoning with human-like cognitive patterns. \Gls{tr} and \Gls{tl} impact \gls{llm} reasoning, with \gls{tr} performing well at small scales and \gls{tl} improving with larger models~\citep{pan_what_2023}. Weak \glspl{llm} provides accurate feedback aligning with human preferences, even with reduced computational resources~\citep{shankar_who_2024, zhao2024weak}.
Nonetheless, challenges persist in scaling reasoning fidelity, maintaining long-horizon coherence, and ensuring robustness across diverse, and dynamic environments.

These innovations mark progress, but we argue they often remain bounded by static task assumptions~\citep{2025taletoolaugmented} and pre-configured cognitive scaffolding~\citep{handler2023balancing}. To realize open-ended agency, we advocate that they must be mated to reasoning architectures that evolve in response to social context and allow agents to restructure their internal models through interaction-driven epistemic shifts.

\subsection{Interaction and Communication}

\begin{figure}[htbp]
    \centering
    \includegraphics[width=0.9\linewidth]{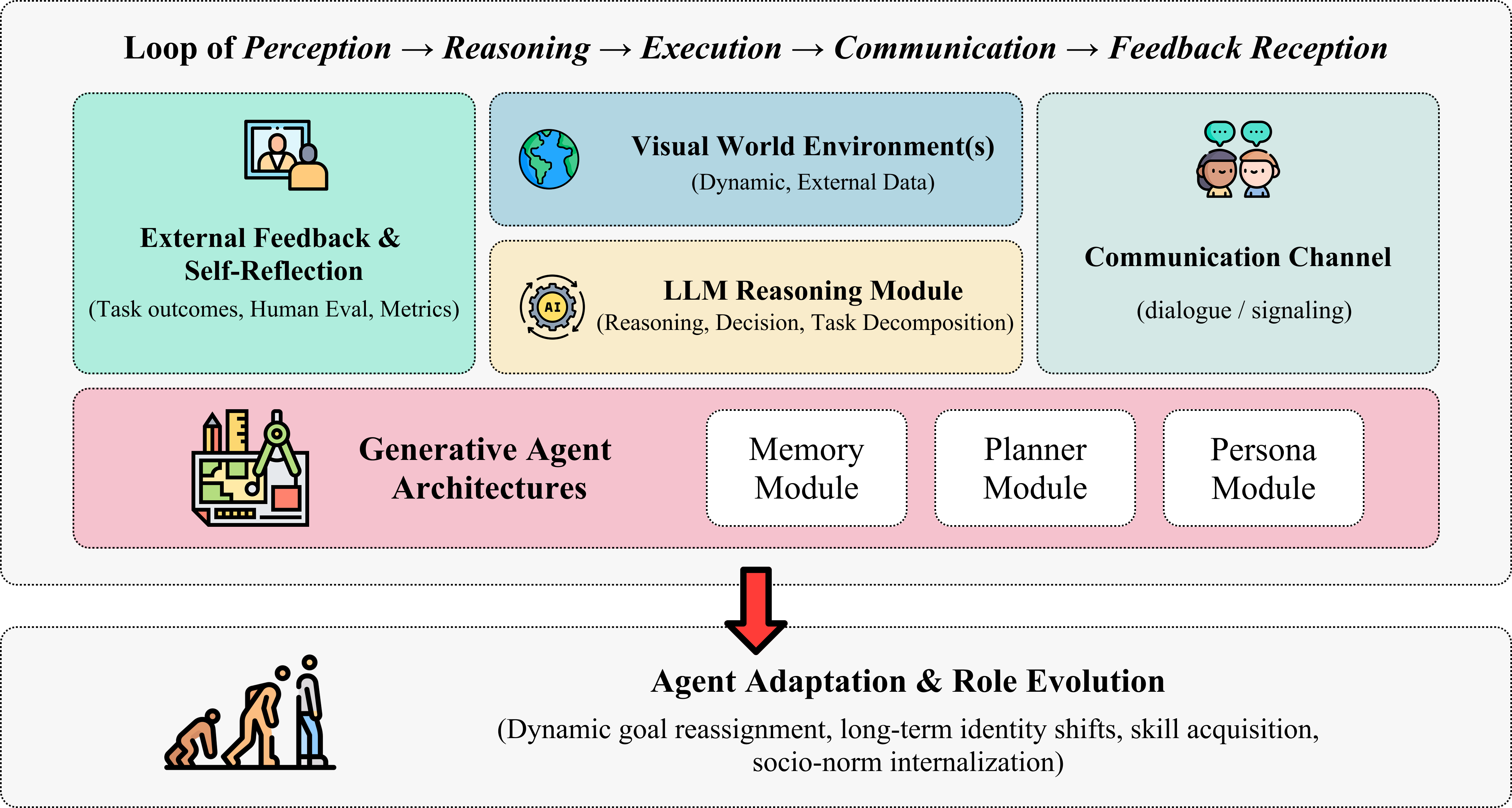}
    \caption{Unified architecture for LLM-driven generative agents in open-ended multi-agent simulations. The upper section depicts the short-term loop: Perception $\rightarrow$ Reasoning $\rightarrow$ Execution $\rightarrow$ Communication $\rightarrow$ Feedback Reception. The lower section highlights long-term development: Agent Adaptation and Role Evolution. Together, these components support both immediate reactivity and sustained co-evolution.}
    \label{fig:architecture}
\end{figure}

Communication in LLM–MAS has evolved from static message passing to richer paradigms of protocol emergence, reflexive dialogue, and relational reasoning. Benchmarks such as ToolQA~\citep{zhuang_toolqa_2023} and inference manipulation strategies~\citep{zhang_model_2024} help optimize knowledge sharing and collaborative planning among agents. Similarly, emerging LLM-based reflexive communication techniques such as response refinement strategies like \gls{rise}~\citep{qu_recursive_2024} and Checkability Training~\citep{kirchner_prover-verifier_2024}, further enhance verifiability and reliability in agent conversations. \gls{rfm}~\citep{tacchetti_relational_2018}. Frameworks like \gls{smart} ~\citep{yue_synergistic_2025} exemplify the growing emphasis on relational reasoning to structure complex multi-agent dialogues.

But although \glspl{llm} have been employed in referential games~\citep{lazaridou_multi-agent_2017, mordatch_emergence_2018} to facilitate the emergence of shared protocols and cooperation strategies using these advances, evaluating emergent communication behaviors at scale and ensuring their alignment with human social norms remain major open challenges~\citep{NEURIPS2024_b35c38f7}. Critically, despite methodological advances, most frameworks treat communication as a means for static task coordination rather than a vehicle for social transformation. We contend that future systems must foreground communication as a dynamic, co-constructive process through which agents evolve new roles, values, and collective norms. A detailed system architecture diagram is provided in~\autoref{fig:architecture}.

\section{Generative Agents for Open-Ended Social Simulation}
\label{sec:generative-agents}

\textbf{Generative agents must move beyond lifelike behavior and toward long-term social transformation.} Moving beyond individual agent capabilities, generative agents powered by \glspl{llm} enable simulations that capture human societies' complexity, adaptability, and emergent dynamics. To highlight the necessity of this approach, the current section critically examines relevant frameworks and methodologies for deploying generative agents in social simulations, identifying key approaches for fostering adaptive behaviors and emergent social phenomena.

\subsection{Frameworks for Generative Agents in Social Simulations}
Frameworks for generative agents seek to create dynamic, scalable environments that take advantage of their ability to demonstrate lifelike adaptability and continuous evolution. Systems such as IICA~\citep{noauthor_framework_nodate} leverage statistical calibration for more robust social simulations, while frameworks like GAA~\citep{yu_affordable_2024} and AGA~\citep{park_generative_2024} integrate \glspl{llm} to support real-time responses and policy reuse, respectively. In concert, these advances mark a shift away from rigid, rule-based models toward fluid, context-sensitive interactions.

Other recent developments combine \gls{marl} with \gls{gabm}~\citep{dizaji_incentives_2024} to simulate more nuanced social behaviors. Deploying \glspl{llm} directly on devices~\citep{zou_wireless_2023} further enhances agent responsiveness in distributed environments by reducing latency. Notably, frameworks such as \gls{smart}~\citep{sato_gai_2024} and SurrealDriver~\citep{jin_surrealdriver_2024} emphasize growing priorities around interpretability, memory integration, and emulating human-like decision-making processes. 

But while these frameworks demonstrate flexibility and responsiveness, their adaptability is often bounded by pre-design objectives. Fully living up to their potential and true alignment with real-world needs requires future systems to move beyond such scenario-specific tuning toward agents capable of revising their social roles, negotiating novel norms, and engaging in value-driven transformation across diverse environments.~\textbf{We contend that generative agents must continuously refine and reconfigure their social roles and behaviors in response to dynamic environments and interactions, rather than merely optimizing for predefined goals.}

\subsection{Methodologies for Simulating Social Interactions}
Improving methodologies for simulating social interactions can be categorized into three target dimensions: Adaptive Behaviors, Cognitive Processes, and Communication and Reasoning, and collectively contribute to advancing the fidelity and richness of social simulations. For a structured summary of simulating social interactions with LLM-driven generative agents, see Table~\ref{tab:social_methods} in Appendix~\ref{appendix-a}.

Rigorous social simulation requires agents capable of nuanced, context-aware interactions. Methodologies like Reflexion~\citep{shinn_reflexion_2023} and Reflection modules~\citep{feng_role_2023} empower agents to modulate their behaviors based on past experiences, while hierarchical role-playing~\citep{deng_are_2025, shao_character-llm_2023} facilitates the representation of intricate social roles and relational dynamics. Cognitive strategies, including \gls{cgcot} prompting~\citep{wu_concept-guided_2024} and task decomposition~\citep{kang_researcharena_2025}, augment agents' planning and problem-solving capabilities, and systems such as Lyfe Agents further refine multi-stage cognitive reasoning within fluid social environments~\citep{kaiya_lyfe_2023}. 

Regarding communicative aspects, referential games~\citep{lazaridou_multi-agent_2017, mordatch_emergence_2018} foster the evolution of shared linguistic conventions and coordination paradigms, whereas advancements in causal reasoning~\citep{imai_causal_2024} and auditing methodologies~\citep{jiao_navigating_2025}  bolster the transparency, verifiability, and safety characteristics of agent interactions. See Appendix~\ref{sec:appendix-b} for representative task scenarios enabled by these frameworks. However, although these methodologies significantly improve social simulation fidelity, systems applying them often operate within pre-defined cultural or task assumptions. It is our position that future agent modeling necessitates the inclusion of not only interactional realism but also cross-contextual generalization and the ability to derive novel social ontologies through bottom-up processes. \textbf{Future methodologies should prioritize emergent norm formation and open-ended adaptation across diverse contexts.}

\subsection{Adaptive Behaviors and Emergent Phenomena}
The development of generative agents capable of modeling complex social systems necessitates the fostering of adaptive behavior and emergent phenomena.

Currently, frameworks such as Richelieu~\citep{guan_richelieu_2024}, StateAct~\citep{rozanov_stateact_2025}, and EABSS~\citep{siebers_exploring_2024} facilitate iterative skill development, self-evolution, and the modeling of complex behaviors. Memory architectures, such as DHMA~\citep{hou_my_2024}, ensure coherent multi-turn interactions, thereby enabling agents to maintain consistent personas and social trajectories over extended periods. Furthermore, initiatives such as SAPLMA~\citep{azaria_internal_2023} integrate internal state analysis and trust metrics, both of which are critical for the construction of reliable and socially acceptable agents. Structured feedback mechanisms, exemplified by OCRM~\citep{sun_rethinking_2025}, further drive strategy refinement, meta-cognitive self-improvement, and the emergence of novel social behaviors over time.

Despite meaningful progress in modeling local social interactions, current generative social modeling frameworks often fall short in enabling systemic societal evolution. Many simulations still overfit to bounded task spaces or culturally narrow contexts, limiting their broader applicability. It is important that future research prioritize creating agents capable of cross-contextual adaptation and sustained societal innovation beyond initial design constraints. This paper thus advocates for research that explicitly models long-range normative drift, inter-agent value negotiation, and open-ended co-creation of social institutions.

\section{Open-ended Simulation and Co-evolution}

To enhance the capacity of \gls{llm} driven simulations, open-ended simulations are designed to represent the ongoing development of agents and environments. This allows for the analysis of intricate adaptive systems, emergent occurrences, and fluid social frameworks. In contrast to conventional task-oriented configurations, open-ended simulations prioritize sustained innovation, reciprocal modification, and the manifestation of unanticipated behaviors over extended durations.~\textbf{But despite the important strides existing work has made, constraints remain due to rigid environmental scaffolds and narrow adaptation loops. To move forward, there is a need to build systems designed for long-term transformation, not just short-term task generalization.} This segment seeks to further substantiate the paper's central premise that the progression of adaptive Artificial Intelligence (AI) is contingent upon the adoption of simulations where agents and environments undergo simultaneous evolution in limitless and unforeseen manners.

\subsection{Dynamic Scenario Evolution}
Dynamic scenario evolution refers to simulations where environments change continuously, driven by both agent interactions and external inputs. Frameworks such as VOYAGER~\citep{wang_voyager_2023} and OpenWebVoyager~\citep{he_openwebvoyager_2024} demonstrate that agents can autonomously explore, adapt, and learn in evolving worlds. By engaging in iterative feedback loops, agents refine their strategies and progressively build skill repertoires to navigate ever-changing environments. For example, in financial simulations, dynamic adaptation mechanisms inspired by human memory systems allow agents to adjust flexibly to shifting market conditions~\citep{li_tradinggpt_2023}. Similarly, metacognitive modules~\citep{toy_metacognition_2024} use introspection to enhance agents' ability to self-monitor and refine decision-making across increasingly complex tasks, supporting sustained autonomy over time. 

Meanwhile, incorporating real-world data, such as public sentiment in opinion dynamics simulations~\citep{wei_mimicking_2024}, allows agents to adapt their behavior, while diversity-driven data augmentation~\citep{sahu_promptmix_2023} enhances scenario complexity to reflect real-world unpredictability. However, current dynamic scenario evolution techniques often rely on predefined rule adjustments or superficial environmental changes, which limit their overall potential. 


In contrast, this paper advocates for simulation environments that treat unpredictability as a design objective, where agents are evaluated not just on task success but on their capacity to generate, reshape, and outgrow task definitions themselves.~\textbf{Truly open-ended systems must empower agents to co-create new task domains, communication protocols, and value systems.}

\subsection{Co-evolution of Agents and Environments}
Co-evolution highlights the potential for reciprocal adaptation between agents and their environments. As agents evolve and act, they reshape their surroundings, which in turn influences their subsequent behaviors. This dynamic feedback loop is critical for modeling realistic societal and ecological systems.

Cognitive frameworks such as Global Workspace Theory~\citep{zhang_understanding_2025} allow agents to selectively process environmental information to adapt their behavior. The GABSS framework~\citep{xiao_simulating_2023} illustrates how agents can adjust to public events, fostering emergent phenomena at the societal level. Meanwhile, driving simulations like SurrealDriver~\citep{jin_surrealdriver_2024} show how agent behaviors continuously adapt to human-like driving data, enhancing realism. Similarly, economic models such as TwinMarket~\citep{yang_twinmarket_2025} reveal that micro-level agent decisions aggregate into macro-level market dynamics, emphasizing the importance f modeling agent-environment interplay. Furthermore, marketplace models for AI training~\citep{sarkar_viz_2023} demonstrate that co-evolutionary processes drive innovation, as agents and environments mutually push each other toward novel capabilities.

\textbf{To maximize benefits, we contend that future research must explicitly prioritize mutual transformation between agents and their worlds, rather than merely using environments as static backdrops for agent benchmarking.}

\subsection{Generative Agents in Open-ended Simulations}
Generative agents often act as the engines of adaptability in open-ended simulations. By integrating \glspl{llm} along with movement, perception, and memory systems, agents can interact with dynamic environments in even more sophisticated ways~\citep{verma_generative_2023}. Persona-driven data synthesis techniques~\citep{ge_scaling_2024} further enrich the diversity of integrated agent profiles, enabling more realistic social simulations and synthetic data generation.

Nevertheless, significant challenges persist: hallucinations, alignment issues, and computational costs continue to limit the scalability and reliability of generative agents~\citep{chen_survey_2025, feng_what_2024}. Though stabilization methods, such as concept-based consistency measures~\citep{yang_llm-measure_2024} and external knowledge integration~\citep{wang_t-sciq_2023}, offer promising paths for improving agent reasoning quality. Recent frameworks such as UAG~\citep{ni_towards_2025} aim to equip agents with trustworthy decision-making capabilities in high-stakes contexts. In addition, balancing efficiency, interpretability, and open-ended adaptability remains a critical frontier for future research.



Critically, true co-evolution is not just about accelerating the speed and accuracy with which simulation agents can actuate and respond to environmental change; It requires expanding the breadth and depth of the innovation frontier itself. Agents must be empowered not only to adapt but also to invent new domains of interaction, challenge established norms, and foster emergent complexities that transcend their original operational scope. \textbf{Therefore, this paper calls for a shift from adaptive behavior toward adaptive ontology, where systems empower agents to invent new categories of behavior rather than merely refining those that already exist.} By empowering these innovations with open-ended continuous co-evolution, a future where agents and environments co-evolve in complex, dynamic ways is achievable, supporting the creation of simulations that are more realistic, resilient, and socially meaningful.
 
\section{Emerging Applications and Open Challenges}

The convergence of \glspl{llm} and \gls{mas} is significantly impacting scientific collaboration, societal modeling, and economic simulation, revealing the shortcomings of traditional static simulation. \textbf{We argue that to fully realize the societal value of LLM-driven \gls{mas}, applications must shift from narrow optimization tasks toward fostering open-ended exploration, continuous feedback, and shared knowledge formation.} This section highlights not only emerging applications but also reveals how current deployment trends may reinforce the limitations of static simulation paradigms. 

Three application domains are explored where LLM-driven \gls{mas} are demonstrating growing impact: scientific collaboration, societal modeling, and economic simulation (A detailed review of these applications, including representative systems and open challenges, is provided in Appendix~\ref{sec:appendix-applications}).

\subsection{Outlook and Research Priorities}


A recurring suite of methodological obstacles is evident in nascent applications across various domains. Scalability, interpretability, safety, and cross-cultural adaptability consistently emerge as imperative requisites. Prospective research endeavors should emphasize interdisciplinary cooperation, comprehensive evaluation paradigms, and the judicious incorporation of societal values to effectuate the complete realization of the innovative capacity inherent in large language model-powered multi-agent simulations.

Furthermore, substantial unrealized potential exists for LLM-driven Multi-Agent Systems to revolutionize sectors including education, policymaking, and social innovation, when considering areas outside of present applications. Subsequent investigations ought to actively examine the extent to which open-ended simulations might function not solely as replications of, but as conceptualizations of human societal constructs.

\section{Challenges and Future Directions}


Despite the remarkable progress in integrating \glspl{llm} into \gls{mas} and social simulations, several critical challenges remain. Addressing these challenges is essential for scaling, aligning, and securing adaptive systems while maximizing their societal benefit. In this section, key obstacles are systematically categorized, and research directions for advancing the field are proposed (Specific technical challenges, including memory consolidation, system scaling, and safety risks, are detailed in Appendix~\ref{sec:appendix-implementation}). 

\textbf{Future work must be guided by first principles of open-endedness, co-evolution, and societal embedding, though there are challenges to abandoning retrofitting legacy benchmarks for increasingly dynamic systems.} This section not only lists technical challenges but also articulates why many current research trajectories may remain overly focused on optimization, efficiency, or isolated fixes.


\subsection{Bias, Fairness, and Societal Impact}

LLM-driven agents risk inheriting and amplifying biases present in their training data~\citep{kamruzzaman_woman_2024, chan_mango_2023}, This can result in cultural misinterpretations, the reinforcement of stereotypes, and risks to credibility within simulated environments, even open-ended ones. To move beyond traditional bias mitigation efforts~\citep{zhou_unibias_2024, xu_good_2024}, which are often static benchmarks, future research must prioritize promoting cultural adaptability and inclusivity, ensuring that simulated societies reflect a broad spectrum of values. 

\textbf{Future frameworks must move beyond mitigation toward proactive norm pluralism, ensuring that simulations do not merely reflect but also challenge and reshape social imaginaries.} As AI-driven simulations increasingly influence policymaking and public discourse, developing frameworks for responsible innovation and interdisciplinary governance will become ever more critical.


\subsection{Evaluation of Emergent Behaviors and Open-ended Dynamics}

Currently, existing evaluation benchmarks often prioritize task performance at the expense of capturing emergent system properties~\citep{xiao_simulating_2023, siebers_exploring_2024}. But to fully unlock the potential of open-ended simulations, robust methodologies for evaluating diversity, societal resilience, coordination quality, and long-term innovation must be developed. Proposed evaluation dimensions and metrics are detailed in Appendix~\ref{sec:appendix-c}.

Critically, future ~\textbf{evaluation frameworks should not merely reward expected outcomes, but also emphasize the discovery and characterization of novel, unanticipated behaviors.}


\subsection{Open-ended and Co-evolutionary Simulation Design}

The development of genuinely open-ended, co-evolutionary simulations necessitates the concurrent evolution of agents and environments, fostering a continuous cycle of challenge and adaptation~\citep{wang_voyager_2023, verma_generative_2023}. Realization of this objective requires advancements in dynamic environment generation, evolutionary agent architectures, and adaptive trust calibration. System designs must explicitly prioritize novelty and resilience, embracing inherent unpredictability over the imposition of constraints.

\subsection{Research Priorities and Roadmap}

\textbf{We predict that within the next decade, open-ended, co-evolutionary multi-agent simulations are poised to become the leading testbed for adaptive AI, overtaking static benchmarks. However, realizing this potential necessitates a fundamental shift in how we approach evaluation standards, safety protocols, and societal modeling. }

Drawing on identified challenges, the following are key priorities for future research: (a) Designing culturally adaptive, memory-augmented generative agents capable of lifelong learning and dynamic role adaptation. (b) Developing lightweight, communication-efficient \gls{llm} surrogates that enable scalable, heterogeneous simulations. (c) Formalizing evaluation frameworks for emergent behaviors, systemic resilience, and societal alignment. (d) Embedding continuous safety monitoring, fairness auditing, and ethical safeguards into multi-agent architectures. (e) Fostering interdisciplinary collaboration among AI researchers, cognitive scientists, ethicists, and policymakers to align simulation objectives with broader societal values.


By embracing unpredictability and focusing on long-term adaptability, the research community can move closer to building resilient, interpretable, and socially beneficial LLM-driven multi-agent ecosystems. By examining these areas, a more profound understanding of societal complexity, adaptation, and innovation can be achieved. This deeper knowledge can then pave the way for a new generation of adaptable and continuously evolving AI systems.

\section{Conclusion}

This research critically examines prevailing assumptions concerning the integration of \glspl{llm} within \gls{mas} and social simulations. While acknowledging certain constrained successes, extant systems exhibit a deficiency in accurately reflecting the dynamic, progressive, and complex social attributes of real-world societies. In addressing these limitations, this paper proposes a transition from static, objective-oriented assessments to a taxonomy grounded in fluctuating situational dynamics, the co-evolution of agents and their environments, and the deployment of generative agent architectures.

Further, it offers a research roadmap outlining social fairness, evaluation and design challenges needing addressing, including long-term memory consolidation, bias mitigation, scalable efficiency, and emergent behavior evaluation. These priorities reflect the broader aim to replace brittle optimization pipelines with adaptable, resilient, and norm-aware agent ecosystems.

\textbf{Looking ahead, we contend that open-ended, co-evolutionary multi-agent simulations must define the next decade of adaptive AI research. Embracing unpredictability is not a liability to be controlled but a catalyst for innovation, resilience, and societal relevance.}  
\textbf{Only by committing to this shift can we unlock new pathways for fostering emergent innovation, enhancing societal resilience, and advancing the frontier of adaptive AI.}

\backmatter

\bmhead{Acknowledgements}

The authors used a large language model to assist with grammar and stylistic polishing. All substantive ideas, analyses, and conclusions remain the responsibility of the authors, and any suggestions from the tool were reviewed and edited before inclusion.


\bibliography{Multi-agent}




\clearpage

\appendix

\section{Tables for Key Frameworks}
\label{appendix-a}

\begin{table}[htbp]
\centering
\caption{Overview of key frameworks in LLM-driven multi-agent and social simulation
research. Each row summarizes a representative framework by its publication year,
reference source, primary application domain, and key characteristics. This table
highlights the diversity of methods, tasks, and innovation directions within the emerging
field of open-ended multi-agent simulations. Note: Additional simulation techniques and
methodologies are discussed in Section~\ref{sec:generative-agents} but are not fully
represented in this framework table.}
\label{tab:frameworks}
\small
\rotatebox{90}{%
\begin{minipage}{0.92\textheight}
\centering
\begin{tabular}{p{3.0cm}p{0.9cm}p{4.2cm}p{5.8cm}}
\toprule
\textbf{Framework} & \textbf{Year} & \textbf{Application} & \textbf{Key
Characteristics} \\
\midrule
VOYAGER~\citep{wang_voyager_2023} & 2023 & Minecraft open-world exploration &
Automatic task generation, iterative skill acquisition, LLM-driven adaptation \\
OpenWebVoyager~\citep{he_openwebvoyager_2024} & 2024 & Web navigation & Multimodal
exploration, feedback-optimization loops, generalization to unseen domains \\
TwinMarket~\citep{yang_twinmarket_2025} & 2025 & Financial market simulation & Social
influence modeling, private information dynamics, agent heterogeneity \\
AI-Economist~\citep{dizaji_incentives_2024} & 2024 & Economic policy optimization &
Reinforcement learning for tax policy, social welfare maximization \\
\gls{gabm}~\citep{dizaji_incentives_2024} & 2024 & Societal simulations under varied
governance & Skill trade modeling, institution-driven agent behavior evolution \\
SurrealDriver~\citep{jin_surrealdriver_2024} & 2024 & Driving simulation & LLM-based
driver agent modeling, naturalistic trajectory generation \\
\gls{smart}~\citep{yue_synergistic_2025} & 2025 & Knowledge-intensive task
collaboration & Trajectory learning, multi-agent relational reasoning \\
Richelieu~\citep{guan_richelieu_2024} & 2024 & AI diplomacy and negotiation & Self-
evolving agents, skill co-adaptation \\
StateAct~\citep{rozanov_stateact_2025} & 2025 & Long-term adaptation simulation &
State-tracking, dynamic role evolution \\
UAG~\citep{ni_towards_2025} & 2025 & Trustworthy graph reasoning & Uncertainty-aware
decision making, reasoning under ambiguity \\
EABSS~\citep{siebers_exploring_2024} & 2024 & Social simulation model design &
Conversational AI-supported conceptual modeling, stakeholder-driven co-creation \\
GAA~\citep{yu_affordable_2024} & 2024 & Cost-efficient agent simulations & Policy
reuse, social memory compression \\
AGA~\citep{park_generative_2024} & 2024 & Mass-scale human behavior simulation &
Personality-grounded generative agents, large-scale validation \\
IICA~\citep{noauthor_framework_nodate} & 2013 & Social simulation & Statistical
calibration, indirect inference, Gaussian mixture modeling \\
\bottomrule
\end{tabular}
\end{minipage}}
\end{table}


\begin{table}[htbp]
\centering
\caption{Key methodologies supporting generative agents and social simulations. This table complements the framework overview by summarizing essential methods that enhance agent adaptability, cognition, and communication. Each method is categorized based on its primary function in agent simulation pipelines.}
\label{tab:social_methods}
\small
\rotatebox{90}{%
\begin{minipage}{0.88\textheight}
\centering
\begin{tabular}{p{3.0cm}p{2.6cm}p{4.2cm}p{5.8cm}}
\toprule
\textbf{Method} & \textbf{Category} & \textbf{Function} & \textbf{Key Characteristics} \\
\midrule
Reflexion~\citep{shinn_reflexion_2023} & Adaptive Behavior & Behavioral refinement & Promotes iterative self-improvement via feedback and memory recall \\
Self-Evaluation Module~\citep{feng_role_2023} & Adaptive Behavior & Meta-cognitive adaptation & Enables reflective planning based on past performance \\
Hierarchical Role-Playing~\citep{deng_are_2025} & Adaptive Behavior & Role-specific reasoning & Supports multi-level identity simulation and narrative consistency \\
\gls{cgcot} Prompting~\citep{wu_concept-guided_2024} & Cognitive Process & Chain-of-thought refinement & Leverages concept-driven reasoning for strategic planning \\
Task Decomposition~\citep{kang_researcharena_2025} & Cognitive Process & Planning and modularization & Breaks complex goals into manageable sub-goals \\
Lyfe Agents~\citep{kaiya_lyfe_2023} & Cognitive Process & Dynamic reasoning & Integrates multi-stage memory and event-triggered logic \\
Referential Games~\citep{lazaridou_multi-agent_2017} & Communication & Protocol emergence & Fosters shared language and cooperative strategies \\
Causal Reasoning~\citep{imai_causal_2024} & Communication & Logic and attribution & Enhances verifiability through explainable cause-effect inference \\
Auditing Methodologies~\citep{jiao_navigating_2025} & Communication & Transparency enforcement & Structures traceable, value-aligned decision-making pathways \\
\bottomrule
\end{tabular}
\end{minipage}}
\end{table}


\clearpage

\section{Application Case Studies}
\label{sec:appendix-applications}

\subsection{Multi-Agent Decision-Making and Human-AI Collaboration}

Large language model-based multi-agent systems are significantly altering decision-making processes in complex and dynamic environments. Illustrative examples, such as ResearchArena~\citep{kang_researcharena_2025} and ResearchAgent~\citep{chen_survey_2025} demonstrate the potential to automate literature reviews and enhance hypotheses. Such synergistic intelligence offers great potential for expediting scientific advancement.


In robotics, systems like BrainBody-LLM~\citep{zhang_large_2023} and scalable skill learning frameworks~\citep{ha_scaling_2023} utilize natural language-driven planning to bridge the gap between human intent and robotic execution. Zero-shot human modeling~\citep{mishra_human-mediated_2024} further empowers agents to anticipate and adapt to human preferences with minimal prior training.

Healthcare and safety-critical domains are also seeing significant benefits. LLM-enhanced systems show capacity to improve diagnosis (e.g., SRLM~\citep{wang_unifying_2024}), clinical communication (e.g., MedAlign~\citep{fleming_medalign_2023}), and safety assurance (e.g., Aegis~\citep{ghosh_aegis_2024}).

Nevertheless, future work must address the reliability of multi-agent collaboration under uncertainty, the ethical implications of human-AI partnerships, and the transparency of joint decision-making processes.

In addition, future empirical research should design comparative evaluations that treat \gls{llm} collectives as potential epistemic partners, or even challengers, to human decision-making norms, rather than mere as tools.

Future systems must treat \gls{llm} collectives not merely as productivity tools, but as epistemically dynamic collaborators capable of revising shared knowledge, ethical norms, and strategic goals over time.

\subsection{Simulation for Societal Systems and Urban Environments}

\glspl{llm} are creating novel opportunities for simulating societal structures, urban dynamics, and the effects of policies. For instance, tools like ChatMap~\citep{unlu_chatmap_2023} and GrutopiaDream~\citep{zhou_large_2024} allow for city modeling through natural language, thereby increasing public involvement in urban planning.

Generative agent models facilitate the analysis of human-environment interactions~\citep{verma_generative_2023}, providing enhanced understanding of social behavior, environmental perception, and the modeling of public responses. Similarly, simulations of opinion dynamics~\citep{wei_mimicking_2024} can help illustrate the evolution of collective sentiment in reaction to dynamic societal factors.

Recent work in this field increasingly emphasizes emotional adaptation~\citep{meyer_you_2024}, cross-cultural modeling~\citep{kamruzzaman_woman_2024}, and realistic conversational norms~\citep{chan_mango_2023, lee_thanos_2024} in a bid to enhance the fidelity of social simulations.

However, enhancing the interpretability of emergent societal phenomena, mitigating cultural biases, and ensuring fairness across diverse populations remain major open challenges.

Regardless, we contend that simulations have to evolve from static scenario rendering toward living, feedback-driven platforms where public values, emotional dynamics, and environmental responses co-develop in real time.

\subsection{Economic Behavior, Finance, and Cultural Dynamics}

In the fields of economics and finance, \glspl{llm} facilitate sophisticated simulations of decision-making processes characterized by uncertainty and asymmetric information. Notable examples such as AI-Economist~\citep{dizaji_incentives_2024} and TwinMarket~\citep{yang_twinmarket_2025} illustrate the capacity of multi-agent simulations to model complex phenomena, including negotiation, market formation, and social influence.

Efficiency-oriented techniques like TrustScore~\citep{zheng_trustscore_2024} help facilitate scalable simulations by balancing resource consumption and fidelity. In parallel, cultural adaptation frameworks explore how agents internalize and operate across diverse social norms and negotiation styles.

However, key challenges in this area include preserving behavioral diversity at scale, accurately modeling macroeconomic emergence from micro-level interactions, and evaluating the systemic effects of adaptive agent behaviors.


\clearpage

\section{Representative Scenarios}
\label{sec:appendix-b}

This appendix provides a set of representative task scenarios for the frameworks summarized in Appendix A. Each scenario highlights how different systems operationalize open-ended behaviors, societal adaptation, and domain-specific innovation.

\begin{itemize}
    \item \textbf{VOYAGER}: Agents autonomously discover and build technologies in a Minecraft world, achieving faster exploration milestones compared to task-fixed agents.
    \item \textbf{OpenWebVoyager}: Agents iteratively improve web browsing capabilities, autonomously learning to navigate and interact across real-world websites.
    \item \textbf{TwinMarket}: Agents in financial markets adapt their trading behavior based on peer influence and private information, simulating realistic market dynamics.
    \item \textbf{AI-Economist}: Agents co-learn economic policies that balance productivity and equality through reinforcement learning in simulated societies.
    \item \textbf{\gls{gabm}}: Agents evolve their skillsets and economic behaviors based on incentive structures in varied virtual governance environments.
    \item \textbf{SurrealDriver}: LLM-based agents emulate human-like driving patterns, adapting to diverse driving styles and environmental changes.
    \item \textbf{\gls{smart}}: Agents collaborate on complex knowledge tasks by learning from past solution trajectories and relational reasoning.
    \item \textbf{Richelieu}: Agents negotiate and evolve diplomatically, developing complex strategies over time in evolving political landscapes.
    \item \textbf{StateAct}: Agents improve long-term adaptability by dynamically modifying their internal states and behavioral strategies.
    \item \textbf{UAG}: Agents reason under uncertainty within knowledge graphs, improving trust and robustness in decision-making.
    \item \textbf{EABSS}: Conversational AI systems assist in the co-creation of social simulation models with domain stakeholders.
    \item \textbf{GAA}: Agents interact with the environment and peers cost-effectively by learning lifestyle policies and social compression techniques.
    \item \textbf{AGA}: Over 1,000 real individuals are simulated using personality-grounded generative agents, enabling social science experiment replication.
\end{itemize}

\clearpage

\section{Evaluation Metrics for Open-Ended Multi-Agent Simulations}
\label{sec:appendix-c}

This appendix outlines proposed evaluation dimensions and metrics to assess the richness, adaptability, and societal relevance of open-ended multi-agent simulations. These metrics are intended to complement traditional task-based evaluations by emphasizing emergent phenomena and long-term innovation.

To properly evaluate open-ended multi-agent simulations, we propose the use of the following key dimensions and example metrics:

\begin{itemize}
    \item \textbf{Exploration Capacity}:
    \begin{itemize}
        \item Unique state/item discovery count
        \item Coverage rate across environment features
        
        Shannon entropy: \( H = -\sum p(x) \log p(x) \)
    \end{itemize}
    \item \textbf{Adaptation and Learning}:
    \begin{itemize}
        \item Task success improvement over time
        \item Strategy change frequency under environmental shifts
    \end{itemize}
    \item \textbf{Emergent Innovation}:
    \begin{itemize}
        \item Novel behavior emergence rate
        \item Diversity of discovered social norms or artifacts
    \end{itemize}
    \item \textbf{Societal Stability and Alignment}:
    \begin{itemize}
        \item Resilience to agent/environment perturbations
        \item Alignment with externally defined social value norms
    \end{itemize}
    \item \textbf{Memory and Consistency}:
    \begin{itemize}
        \item Longitudinal consistency score
        \item Memory fidelity over multiple interaction rounds
    \end{itemize}
\end{itemize}

These metrics aim to capture not only task performance but the richness, novelty, and social realism of open-ended, evolving agent societies.

\clearpage

\section{Architectural and Operational Challenges}
\label{sec:appendix-implementation}

\subsection{Memory Consolidation and Temporal Reasoning}

Autonomous agents frequently exhibit difficulty in sustaining coherent long-term memory and robust temporal reasoning within changing contexts ~\citep{hou_my_2024, toy_metacognition_2024}. While architectures such as DHMA~\citep{hou_my_2024} present encouraging advancements, persistent challenges are evident in continuous learning, catastrophic forgetting, and the dynamic updating of knowledge. Integrating metacognitive modules, which allow agents to observe and refine their memory systems, has the potential to significantly improve adaptability over extended periods.

Agents should conceptualize memory not solely as a repository, but as a dynamic framework that supports identity continuity, social responsibility, and epistemological development.

\subsection{Scaling, Efficiency, and Diversity Preservation}

The implementation of LLM-driven systems for large-scale, heterogeneous multi-agent simulations presents substantial resource challenges~\citep{yao_nanolm_2024, yang_predictive_2024}. Although strategies such as model distillation, adaptive prompting, and predictive evaluation~\citep{zheng_trustscore_2024} offer partial reductions in computational expenditures, they frequently compromise the retention of behavioral diversity. Future investigations should prioritize the development of communication-efficient architectures, memory compression methodologies, and decentralized learning paradigms to sustain agent heterogeneity while simultaneously ensuring scalability.

Rather than treating scale and diversity as trade-offs, we call for research that designs architectures where heterogeneity and generalization reinforce rather than oppose each other.

\subsection{Security, Safety, and Ethical Considerations}

With the growing autonomy and pervasiveness of adaptive agents, concerns regarding security vulnerabilities, the spread of misinformation, and potential dual-use applications are amplifying~\citep{feng_what_2024, ghosh_aegis_2024}. It is imperative to integrate verifiable reasoning, transparent decision-making processes, and mechanisms that ensure privacy preservation into the architectural design of these agents~\citep{jiang_identifying_2023, shi_know_2025}. Promising models for the responsible and secure implementation of agents include frameworks such as Aegis~\citep{ghosh_aegis_2024} and continuous ethical audits~\citep{jiao_navigating_2025}.




\end{document}